\documentclass[12pt]{iopart}

\usepackage{iopams}
\usepackage{graphicx}
\usepackage{float}

\begin{document}

\title[ ]{Polarization attack on continuous-variable quantum key distribution}

\author{Yijia Zhao $^1$, Yichen Zhang$^{1}$, Yundi Huang $^1$, Bingjie Xu$^2$, Song Yu$^1$, Hong Guo$^3$}

\address{$^1$State Key Laboratory of Information Photonics and Optical Communications, Beijing University of Posts and Telecommunications, Beijing 100876, China}

\address{$^2$Science and Technology on Security Communication Laboratory, Institute of Southwestern Communication, Chengdu 610041, China}

\address{$^3$State Key Laboratory of Advanced Optical Communication System and Network, School of Electronics Engineering and Computer Science and Center for Quantum Information Technology, Peking University, Beijing 100871, China}
\ead{yusong@bupt.edu.cn and hongguo@pku.edu.cn}

\begin{abstract}
The shot-noise unit (SNU) is a crucial factor for the practical security of a continuous-variable quantum key distribution system. In the most widely used experimental scheme, the SNU should be calibrated first and acts as a constant during the key distribution. However, the SNU of a practical system is dependent on the various parameters of the local oscillator, which can be controlled by the eavesdropper in the open channel. In this paper, we report a quantum hacking method to control the practical SNU by using the limited compensation rate of the polarization compensation. Since the compensation is only based on of the polarization measurement results of part of local oscillator pulses, the polarization of other unmeasured pulses may not be compensated correctly, which can be utilized by the eavesdropper to control the practical SNU. The simulation and experiment results indicate that the practical SNU can be controlled by the eavesdropper. Thus, the eavesdropper can use the fact that the practical SNU is no longer equals to the calibrated one to control the excess noise and final key rate.
\end{abstract}

\pacs{03.67.Dd, 03.67.Hk}
\submitto{\JPB}
\maketitle

\section{Introduction}
Continuous-variable quantum key distribution (CV-QKD) ensures that two legal parties(Alice and Bob) generate secure keys through an untrusted channel~\cite{CV3,CV4}. The physical implementation of CV-QKD using Gaussian modulated coherent state is based on optical communication techniques which are beneficial to a wide range of practical applications~\cite{GMCS1,GMCS2,NETWORK}. With the development of CV-QKD experiment~\cite{EXPERIMENT5,EXPERIMENT51,EXPERIMENT1,EXPERIMENT2,EXPERIMENT23}, the transmission distance of CV-QKD system in the field test has reached 50km, which is enough to support the construction of metropolitan networks~\cite{EXPERIMENT22,EXPERIMENT4}. Meanwhile, CV-QKD protocols based on Gaussian modulated coherent state have been proved to be theoretically secure against general attacks both in the asymptotic case~\cite{theorysecurity1,theorysecurity2,theorysecurity3} and the finite-size regime~\cite{theorysecurity4,theorysecurity5}. The recently proposed user-defined QKD~\cite{udq} that allows one to securely construct the protocol using arbitrary non-orthogonal states, thus will promote the security and implementation of discrete-modulated CV-QKD.

Although the theoretical security of CV-QKD protocols has been proved, many assumptions are made in the process of security analysis. However, the imperfect linearity of the homodyne detector and the transmittance of the beam splitter can be utilized by the eavesdropper to implement saturation attack~\cite{sat,sat1} and wavelength attacks~\cite{wav1,wav2}. The shot-noise unit (SNU) should be constant and act as a normalization parameter of quadrature measurement results~\cite{CV3}. The calibrated SNU is measured by Bob via the interference between the local oscillator (LO) and the vacuum mode before the key distribution. Because the quadrature measurement results of Bob is normalized by the calibrated SNU but scaled with the practical SNU, the consistency of the calibrated and the practical SNU is crucial for the security of a CV-QKD system. The LO is a necessary auxiliary light beam for homodyne detection and the calibration of the SNU, which is assumed not to be tampered by eavesdropper~\cite{xma}. As a result, eavesdropper can bias the practical SNU by manipulating the LO in the channel when CV-QKD system is running~\cite{CVattack2,CVattack1,CVattack3}. In the existing quantum hacking scheme, the eavesdropper changes the practical SNU by controlling the delay between the LO and the sampling clock~\cite{CVattack2} or the intensity of LO~\cite{CVattack1}, so as to reduce the estimation result of excess noise.

In this paper, we propose a polarization attack method against the CV-QKD system by manipulating the polarization of the LO pulses during key distribution. Because the polarization drift rate is usually slower than the repetition frequency of the CV-QKD system, the polarization compensation in a practical CV-QKD system is implemented by measuring a part of LO pulses, which defined as reference LO pulses here. Although the physical implementation is simplified in this way, unmeasured LO pulses can be easily manipulated by the eavesdropper to control the practical SNU. The eavesdropper can make the calibrated and practical SNU unequal by applying the polarization attack to hide the introduced excess noise.

The paper is organized as followed: In section~\ref{sec:2}, we explain the polarization attack method in detail. The eavesdropper can arbitrarily change the practical SNU by applying specific polarization modulation on the unmeasured LO pulses. In section~\ref{sec:3}, the estimated transmittance and excess noise under this attack are given. Alice and Bob will incorrectly estimate the excess noise and a security loophole reveals when the practical SNU is not equivalent to the calibrated value. In section~\ref{sec:4}, numerical simulation results demonstrate that the secret key rate will be overestimated by Alice and Bob when the practical SNU is controlled by Eve. The conclusion and the countermeasure against the polarization attack are discussed in section~\ref{sec:5}.

\section{\label{sec:2} Control the practical shot-noise unit by the polarization attack}

\begin{figure}[t]
\centering
\includegraphics[width=5.5in]{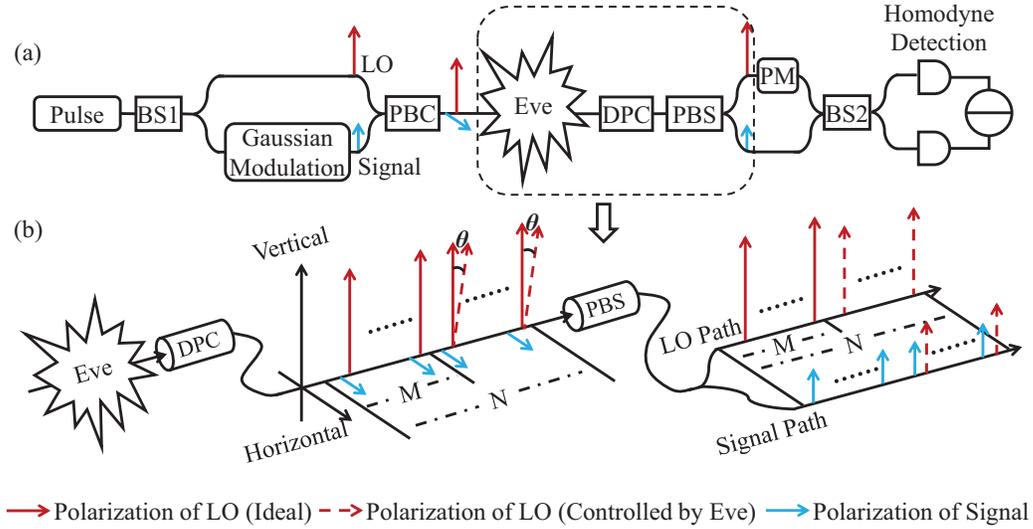}
\caption{ (a) Experimental setup of CV-QKD system~\cite{EXPERIMENT4}. (b) Details of the polarization after the attack. BS1: $1/99$ beam splitter BS2: $50/50$ beam splitter PBC: polarization beam coupler, DPC: dynamic polarization controller, PBS: polarization beam splitter, PM: phase modulation. Homodyne detector~\cite{detector} consists of two photodiodes and a subtractor. Eve controls the orientation angle of the unmeasured LO pulses. The number of reference LO pulses is $M$, the cycle of the polarization compensation is $N$ LO pulses.}
\label{fig1}
\end{figure}

\subsection{Polarization compensation technology}

In current physical implementation of CV-QKD protocols~\cite{EXPERIMENT5,EXPERIMENT1,EXPERIMENT2,EXPERIMENT23,EXPERIMENT22,EXPERIMENT4}, the LO is transmitted together with the signal in one fiber simultaneously to reduce the phase noise generated during the transmission. Time and polarization multiplexing are used in Alice to reduce the crosstalk from the LO (Fig.~\ref{fig1} (a)), while time and polarization demultiplexing are applied by Bob before his homodyne detection. The polarization drift rate of optical pulses in the channel can increase the loss of the signal and the crosstalk from the LO. So polarization compensation is necessary to promote the performance of the system. The current compensation scheme is based on polarization measurement and real-time feedback which depends on the polarization measurement result of the LO.

The polarization compensation implemented in this paper is the general scheme in commercial fiber application. Firstly, a small part of the LO will be divided by a beam-splitter whose polarization state will be measured to calculate the drift from the target polarization state with high precision. Then a feedback signal is generated based on the measured drift and modulated on a polarization controller to compensate the drift.

The polarization of optical pulses in a fiber can be described by the orientation angle $\theta$ and elliptic angle $\varphi$, where $\theta$ and $\varphi$ are used to characterize the orientation and degree of linear polarization respectively. Since the polarization extinction ratio of the laser source in a CV-QKD system is always higher than 30dB (e.g. the laser module from NKT photonics), it is reasonable to assume all optical pulses as linearly polarized. Thus, the elliptic angle of a linearly polarized optical pulse is ${0^ \circ }$, and only the effect of the orientation angle is considered in this paper.

Compared to the high repetition frequency of the system, the polarization drifts slowly enough and can be compensated by only measuring reference LO pulses. For example, the polarization drift rate in the aerial fiber with gale is $34rad/s$, which is several orders of magnitude faster than in the underground fiber~\cite{polarization}. The polarization measurement rate of current commercial devices can reach 2MHz such as EPC1000 (Novoptel) and PSY-201 (General Photonics). If the polarization drift threshold is set as $10^{-5}rad$ per pulse which introduces negligible noise in a system, the necessary compensation rate is just 340kHz. For the CV-QKD system transmitted in the underground fiber, a polarization compensation with kHz rate is efficient enough without any technical barriers. Moreover, the repetition frequency of the current mature CV-QKD system is 5MHz and is still increasing~\cite{EXPERIMENT4}. Thus it's efficient for Bob only to choose a part of the LO pulses in a cycle of the compensation to estimate the drift of the all pulses.

The cycle of compensation is set as $N$ pulses and we consider the polarization difference of those $N$ pulses is within the threshold. Thus, Bob chooses $M$($M \le N$) LO pulses to measure their polarization, which is the reference of the feedback. Eve can determine the reference pulses from all LO pulses by some simple tests. For example, Eve firstly chooses a part of LO pulses to modulate their polarization in one compensation cycle. Because the practical channel parameters (transmittance and excess noise) are controlled by Eve. By comparing the estimated channel parameters, Eve can judge whether these LO pulses are reference pulses. All reference pulses can be determined by Eve performing multiple similar tests.

The LO and signal pulses are split by polarization beam splitters (PBS) into their respective optical path after polarization demultiplexing. If the polarization of the $N-M$ unmeasured LO pulses is not the same as those of the $M$ reference LO pulses, the polarization of this $N-M$ LO pulses cannot be compensated correctly. As shown in Fig.~\ref{fig1} (b), the orientation of $M$ reference LO pulses is parallel to the slow axis of the PBS but there is a misalignment angle $\theta$ between the $N-M$ unmeasured LO pulses and the slow axis. Then the signal pulses will still be splitted into signal path but the LO will also partially leak into the signal path as shown in Fig.~\ref{fig1}. Thus, Eve can control the intensity of $N-M$ unmeasured LO by polarization modulation.

\subsection{Control the practical SNU by polarization modulation on the LO}

Alice selects two sets of random variables X and P from Gaussian distribution random numbers whose variance is $V_A$ and mean value is 0 to prepare Gaussian modulated coherent states. The coherent states and the LO pulses are transmitted to Bob in the same fiber by time-division multiplexing and polarization multiplexing. In the channel, Eve performs collective attack on the coherent states to acquire information of secret keys. The polarization attack is implemented to hide the excess noise introduced from the collective attack. At first, Eve applies some tests to determine the reference pulses. Then, the orientation angle of the unmeasured pulses is modulated by Eve to control the practical SNU. At the receiving end, Bob performs homodyne (X or P) or heterodyne (X and P) detection after polarization compensation and demultiplexing.

Ideally, the slow axis of the PBS is parallel to the orientation angle of all LO pulses and is orthogonal to the orientation angle of all signal pulses after the polarization compensation. We assume that Eve introduces a misalignment angle between the slow axis of the PBS and the orientation angle of the $N-M$ LO pulses. These LO pulses can be decomposed into the parallel part and the orthogonal part. According to the Marius's law, the intensity of the orthogonal part ${I_O}$ which is splitted into the signal path is given by

\begin{equation}
{I_O} = \alpha {I_{LO}}\sin \theta,
\end{equation}
where $I_{LO}$ is the intensity of the LO into the PBS and $\alpha$ is the loss of the PBS. The intensity of the parallel part ${I_P}$ which is splitted into the LO path is

\begin{equation}
{I_P} = \alpha {I_{LO}}\cos \theta.
\end{equation}
According to the above analysis, Eve can control the intensity of the LO for homodyne detection by modulating the orientation angle of the unmeasured LO pulses. The SNU is proportional to the intensity of the LO~\cite{N0LO},

\begin{equation}
{N_0} = \alpha k{I_{LO}}\left\langle {\Delta X_{vac}^2} \right\rangle,
\label{N0}
\end{equation}
where ${N_0}$ is the calibrated SNU, $k$ is a proportional constant dependent on the gain and efficiency of homodyne detector and $\Delta X_{vac}$ is the vacuum fluctuation. As a result, the practical SNU corresponding to the attacked LO pulses is controlled by Eve to

\begin{equation}
{N'_0} = \alpha k{I_{LO}}\left\langle {\Delta X_{vac}^2} \right\rangle \cos \theta.
\end{equation}

\begin{figure}[t]
\centering
\includegraphics[width=4.5in]{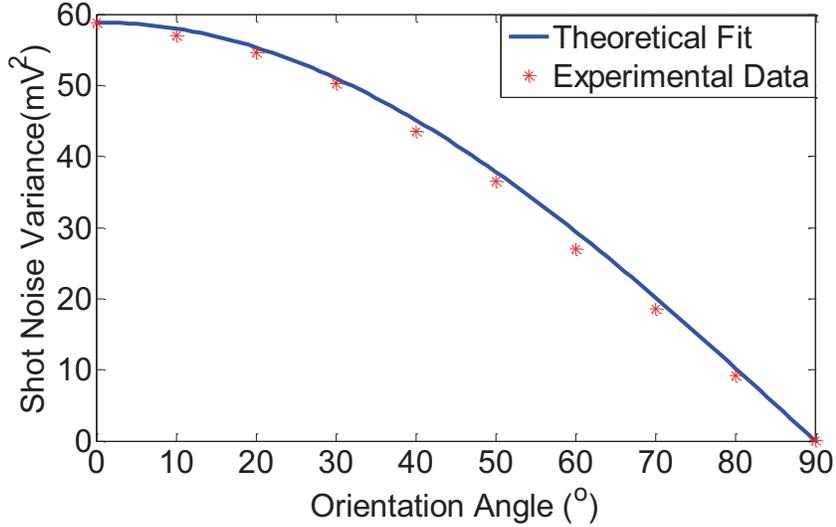}
\caption{ Theoretical simulation shot noise variance (blue solid line) and experimentally measured value (red asterisk) under different orientation angle of LO. }
\label{fig2-1}
\end{figure}

The polarization modulation by Eve may cause the LO to leak into the signal path. The delay between the LO pulses and the signal pulses ensures that the leaked LO will not interfere with the signal. Because the extinction ratio of LO pulses cannot be infinite, there always are some residual photons at the interval of the LO pulses. The polarization modulation will also cause the residual photons to leak to the signal path. Eve can enhance the extinction ratio or not modulate the polarization of the interval to remove the leaked residual photons. The homodyne detection result is normalized by the calibrated SNU which is controlled by Eve under the polarization attack. A simple experiment is implemented to demonstrate the effect of the polarization of the local oscillator on the variance of shot noise. The experimental setup is the receiving end in the Fig1(a). The orientation angle of the LO is modulated from ${0^ \circ }$ to ${90^ \circ }$. The shot noise variance is measured with different orientation angle. Fig.~\ref{fig2-1} shows the experimentally measured shot noise variance with different orientation angle modulation on the LO.

The shot noise variance of ${0^ \circ }$ corresponds to the case without the polarization attack. According to Eq.~\ref{N0}, we plot a theoretical curve which matches well with the experimental result. By modulating the orientation angle of the LO, the practical shot noise variance can be reduced. Therefore, the difference between the calibrated SNU and the practical one is mastered by Eve, which will cause inaccuracy in parameter estimation by Alice and Bob.

\section{\label{sec:3}Practical security analysis under the polarization and collective attack}

The goal of the quantum hacking is to control the results of parameter estimation without being noticed, especially the excess noise. In this section, we demonstrate that Alice and Bob will overestimate the secret key rate under the polarization attack. Based on the current techniques, Eve can implement a kind of non-Gaussian operation which is named as intercept resend attack to obtain secret key information~\cite{intercept}. According to the optimality of Gaussian attack, non-Gaussian attacks will inevitably introduce more excess noise than Gaussian attacks to obtain the same amount of secret key information. It means that Eve must pay more efforts to hide the introduced excess noise. As a result, the security analysis is derived under collective attack which introduces less excess noise to acquire more information of the secret key, and the collective attack can be practically implemented by an entangling cloner attack\cite{entangle1,entangle2}. If the introduced excess noise is not hidden, the information of secret key available for Eve can be estimated and excluded by the privacy amplification~\cite{REVERSE}. In the system with imperfect polarization compensation, the introduced excess noise can be concealed in the parameter estimation by manipulating the orientation angle of the unmeasured LO pulses.

The polarization attack aims at the scheme where the LO and the signal are polarization multiplexed in one fiber. The attack is also valid to other CV-QKD protocols which are implemented with similar scheme, such as ¡°no-switching¡± protocol~\cite{GMCS2}. Since X and P quadratures are symmetrical, we give analysis based on the X quadrature of the homodyne detection for simplicity. The efficiency and electronic noise of the practical homodyne detector are also not considered in this paper. According to the model in the security analysis of optimality of Gaussian attacks~\cite{theorysecurity1,theorysecurity2}, the homodyne detection result ${X_B}$ after normalization without attack is~\cite{normalize}

\begin{equation}
{X_B}{\rm{ = }}\frac{{\left( {\sqrt T {X_A}{\rm{ + }}Z} \right)\sqrt {{N_0}} }}{{\sqrt {{N_0}} }},
\label{XB}
\end{equation}
where ${T}$ is the practical channel transmittance, ${N_0}$ is the calibrated SNU and $Z$ is the total noise. To highlight the effect of the difference between the practical shot noise and calibration value in the parameter estimation, we use the nominator to represent the homodyne detection output before normalization and the denominator to represent the calibration shot noise variance which is used to normalized the detection results. The secret key rate under collective attack is derived from the covariance matrix~\cite{CV3}

\begin{equation}
{\gamma _{AB}} = \left[ {\begin{array}{*{20}{c}}
{\left( {{V_A} + 1} \right){\rm I}}&{\sqrt {T\left( {V_A^2 + 2V_A} \right)} {\sigma _z}}\\
{\sqrt {T\left( {V_A^2 + 2V_A} \right)} {\sigma _z}}&{\left( {T{V_A} + 1 + T\varepsilon } \right){\rm I}}
\end{array}} \right],
\label{CM1}
\end{equation}
where  $\varepsilon$ is the practical excess noise, ${\rm I} = \left[ {\begin{array}{*{20}{c}}
1&0\\
0&1
\end{array}} \right]$ and ${\sigma _z} = \left[ {\begin{array}{*{20}{c}}
1&0\\
0&{ - 1}
\end{array}} \right]$. The transmittance of the channel T and the excess noise $\varepsilon$ are derived with the data of Alice and the homodyne detection result of Bob,

\begin{equation}
\begin{array}{l}
T = {\left( {\frac{{\left\langle {{X_A}{X_B}} \right\rangle }}{{\left\langle {X_A^2} \right\rangle }}} \right)^2}\\
\epsilon = \frac{{\left\langle {X_B^2} \right\rangle  - 1 - T{V_A}}}{T}.
\end{array}
\end{equation}

If Eve applies an orientation angle modulation on the unmeasured LO pulses, the homodyne detection result of the corresponding pulses after normalization reads

\begin{equation}
{X'_B}{\rm{ = }}\frac{{\left( {\sqrt T {X_A}{\rm{ + }}Z} \right)\sqrt {{{N'_0}}} }}{{\sqrt {{N_0}} }} = \sqrt{ \cos \theta} \left( {\sqrt T {X_A}{\rm{ + }}Z} \right).
\label{X'B}
\end{equation}

Similar to the Eq.~\ref{XB}, the numerator indicates the homodyne detection output and the denominator shows the calibrated SNU. Since the practical SNU $N'_0$ is not the same as the calibrated one $N_0$, the normalized measurement results are scaled down by Eve. Accordingly, the estimated transmittance and the excess noise under attack are

\begin{eqnarray}
T' &=& {\left( {\frac{{\left\langle {{X_A}\left[ {\left( {1 - k} \right){{X'}_B} + k{X_B}} \right]} \right\rangle }}{{\left\langle {X_A^2} \right\rangle }}} \right)^2}\\
 \quad \;\; &=& {\left[ {\left( {1 - k} \right)\sqrt {\cos \theta }  + k} \right]^2}T\\
\epsilon' &=& \frac{{\left\langle {{{\left[ {\left( {1 - k} \right){{X'}_B} + k{X_B}} \right]}^2}} \right\rangle  - 1 - T'{V_A}}}{{T'}} \\
 \quad \;\; &=& \epsilon  - \frac{1}{T}\left( {\frac{1}{{{{\left[ {\left( {1 - k} \right)\sqrt {\cos \theta }  + k} \right]}^2}}} - 1} \right),
\label{eqexcess}
\end{eqnarray}
where $k$ is the ratio between the reference LO pulses and the cycle of the compensation which is known as polarization measurement ratio (PMR). The estimated excess noise is dependent both on the PMR $k$ and the orientation angle $\theta$. It is obvious that $\epsilon ' < \epsilon $. The PMR is a constant for a specific system, Eve could reduce the estimated excess noise by modulating the corresponding orientation angle on the unmeasured LO pulses. Thus, she will hide the introduced excess noise as well as obtain the largest amount of secret key information. After the attack, the covariance matrix becomes

\begin{equation}
{\gamma '_{AB}} = \left[ {\begin{array}{*{20}{c}}
{\left( {{V_A} + 1} \right){\rm I}}&{\sqrt {T'\left( {V_A^2 + 2A} \right)} {\sigma _z}}\\
{\sqrt {T'\left( {V_A^2 + 2A} \right)} {\sigma _z}}&{\left( {T'{V_A} + 1 + T'\varepsilon '} \right){\rm I}}
\end{array}} \right].
\label{CM2}
\end{equation}

By comparing the above two covariance matrices Eq.~\ref{CM1} and Eq.~\ref{CM2}, the different elements are the transmittance of the channel and the excess noise which are the decisive parameters for the secret key rate of CV-QKD. Thus, the difference between the practical secret key rate $K\left( {T,\epsilon } \right)$ and the estimated $K\left( {T',\epsilon'} \right)$ under the attack is the amount of the key information which is acquired for Eve. The detailed calculation of secret key $K$ is shown in the appendix.

\section{\label{sec:4} Secret key rate under the collective attack and polarization attack}

Eve's collective attack will introduce excess noise into the Bob's measurement result~\cite{ATTACK}. If Eve doesn't apply the polarization attack, the transmittance and excess noise which are estimated by Alice and Bob will be the practical value $T$ and $\epsilon$. When the polarization attack is implemented to conceal the excess noise, the estimated channel parameters become $T'$ and $\epsilon'$. Since $\epsilon'$ is a function of the practical transmittance and the modulated orientation angle as Eq.~\ref{eqexcess}, Eve needs to adjust the orientation angle of the attack to keep the excess noise within a secure value at different transmission distances. In the following simulation, Alice's modulation variance ${V_A =19}$ and reconciliation efficiency $\beta = 95\%$~\cite{beta,beta1,beta2}.

\begin{figure}[t]
\centering
\includegraphics[width=4.5in]{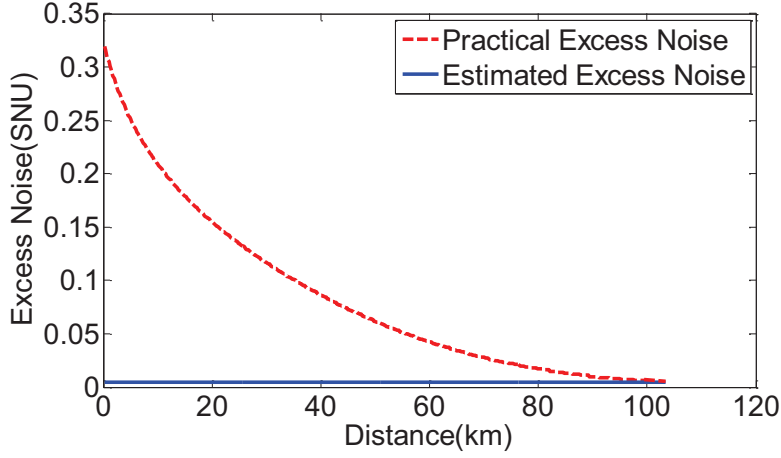}
\caption{ The estimated excess noise under the polarization attack (solid line) and tolerable excess noise under collective attack (dash line). }
\label{fig3}
\end{figure}

The tolerable excess noise is the minimal value that Eve needs to introduce to acquire secret key~\cite{tolerable}. Therefore Eve can acquire all key information by introducing the tolerable noise and reducing the estimated excess noise to a normal value(0.005)~\cite{EXPERIMENT1} by the polarization attack. Fig.~\ref{fig3} shows the minimal introduced excess noise for Eve to acquire all secret key information. By applying the polarization attack, the estimated excess noise can be biased to a normal value 0.005. The PMR $k$ is 0.5 in this simulation.

\begin{figure}[t]
\centering
\includegraphics[width=4.5in]{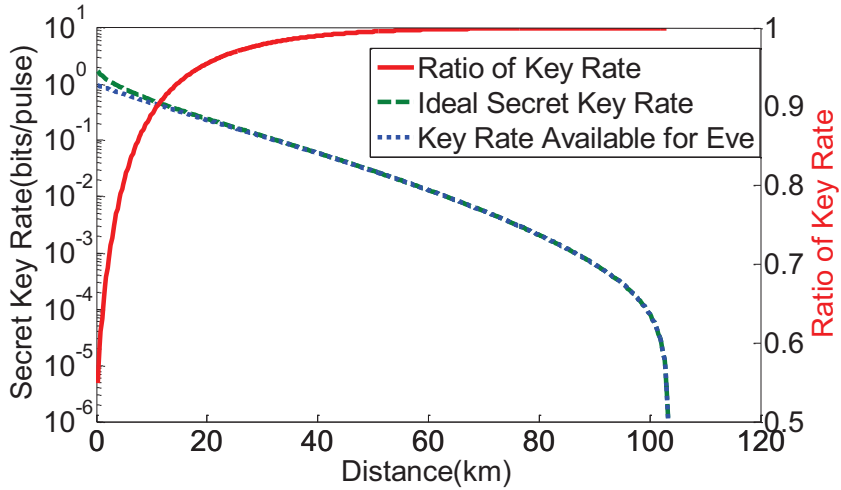}
\caption{ The ideal secret key rate whose the channel parameters are experiment results (dashed line), the maximal secret key rate available for Eve (dotted line) and the ratio of them (solid line, left y axis) vs transmission distance.}
\label{fig4}
\end{figure}

The secret key rate is completely determined by the channel parameters. The secret key rate under the polarization attack or not can be expressed by $K\left( {T',\epsilon'} \right)$ and $K\left( {T,\epsilon } \right)$. We assume that $\epsilon$ introduced by Eve is set as the tolerable excess noise. Therefore, the secret key rate $K\left( {T,\epsilon } \right)$ without the polarization attack remains 0 at all distance and all secret key established by Alice and Bob is insecure. Although the estimated key rate $K\left( {T',\epsilon'} \right)$ seems secure when Eve applies the attack, these secret key information is all available for Eve.

Fig.~\ref{fig4} shows that the attack will slightly reduce the estimated transmittance, so the available key rate $K\left( {T',\epsilon'} \right)$ for Eve will be slightly reduced. The ideal secret key rate $K\left( {T,0.005} \right)$ associates to the channel parameters wihch are experiment results~\cite{EXPERIMENT1} is also illustrated in Fig.~\ref{fig4}. The ratio of $K\left( {T',\epsilon'} \right)$ to $K\left( {T,0.005} \right)$ indicates that the secret key rate available for Eve is gradually approaching to the ideal secret key rate as the distance increases. Because the excess noise needed to be concealed is considerably small as the channel loss increases, the decline of the estimated transmittance caused by the attack is also less significant. Therefore, the polarization attack is more effective and imperceptible with low channel transmission or at long distance ($ \ge 10$km).

\begin{figure}[t]
\centering
\includegraphics[width=4.5in]{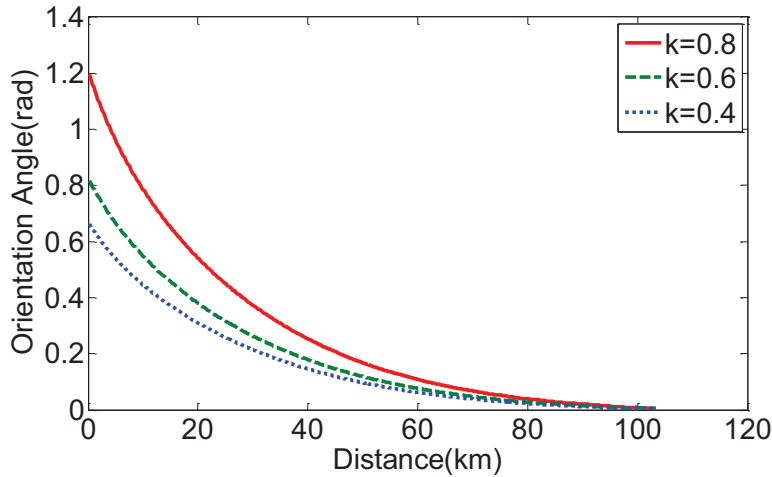}
\caption{ The corresponding orientation angle modulated by Eve vs transmission distance with different $k$: 0.8 (solid line) 0.6 (dashed line) 0.4 (dotted line).}
\label{fig5}
\end{figure}

\begin{figure}[t]
\centering
\includegraphics[width=4.5in]{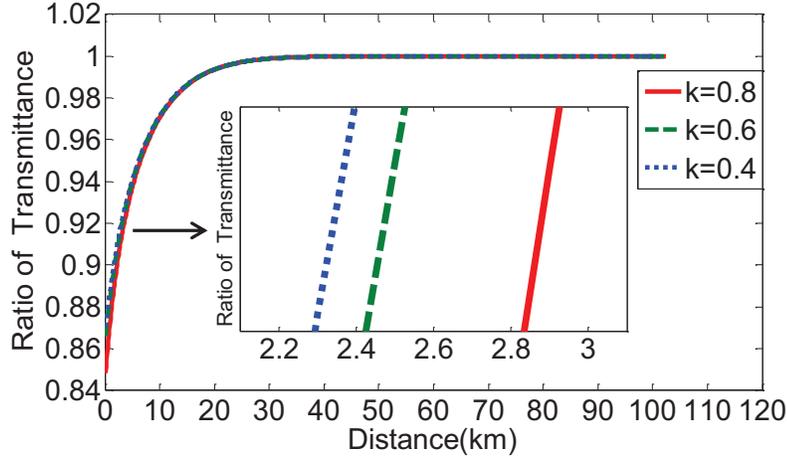}
\caption{ The ratio of the estimated transmittance to the practical value vs transmission distance with $k = 0.8$ (solid line) $k= 0.6$ (dashed line) $k = 0.4$ (dotted line).}
\label{fig6}
\end{figure}

The orientation angle which should be modulated by Eve at different distance is shown in Fig.~\ref{fig5} with different PMR $k$. Because the transmittance $T$ and the minimal introduced excess noise $\epsilon$ both decrease with the increasing of distance, the orientation angle modulated by Eve is also reduced which matches the Eq.~\ref{eqexcess}. Only the unmeasured LO pulses can be manipulated by Eve, the lower PMR suggests that the bigger orientation angle is needed to reduce the estimated excess noise. The influence of the attack is limited by the PMR. The most powerful attack which can be applied by Eve makes $\cos \theta  = 0$, the transmittance and excess noise estimation is reduced to

\begin{equation}
\begin{array}{l}
T' = {k^2}T\\
\varepsilon '{\rm{ = }}\varepsilon  - \frac{1}{T}\left( {\frac{1}{{{k^2}}} - 1} \right).
\end{array}
\end{equation}

As a result, the estimation of legal parties has a minimum value and the value is limited by the PMR. If the PMR is promoted to 100\%, the polarization attack will not be effective to the system.

In realistic environments, the estimated channel transmittance which depends on several factors, for example the phase noise may lead the fluctuation of the estimated transmittance~\cite{normalize}. The transmittance is also affected by the attack. A slight fluctuation of the transmittance is reasonable in CV-QKD, but too much deviation may still be noticed by Alice and Bob. We show the ratio of transmittance between two cases in Fig.~\ref{fig6}. The estimated transmittance is reduced to nearly 85\% of the practical value when the distance is less than 10km. As the distance increases over 30km, the estimated transmittance keeps getting closer to the practical value and they nearly equals. It means that Alice and Bob can not observe the attack by the abnormal transmittance over a certain distance.

\section{\label{sec:5}Conclusion}
We propose a polarization attack against the practical security of CV-QKD systems to conceal the introduced excess noise. Because the polarization drift rate is usually lower than the repetition frequency of the system, the polarization measurement for compensation can be only applied on part of LO pulses to simplify the physical implementation. This scheme opens a practical loophole for Eve to control the homodyne result by modulating the orientation angle of the unmeasured LO pulses. The SNU is positively correlated with the intensity of the LO which can be reduced by the modulation on the orientation angle of the LO pulses in the channel. As a result, the practical SNU can be controlled by Eve during the secret key distribution. The effect of the polarization modulation on LO is also demonstrated by the experiment and the polarization attack can lower the practical SNU than the calibrated one.

The parameter estimation under the polarization attack shows that the estimated excess noise is controlled by Eve with a slight decrease of the estimated transmittance. At different distances, the estimated excess noise can be manipulated at a secure value. Alice and Bob can not observe that the key distribution is no longer secure. Since the attack is based on the limited polarization measurement rate, the most simple countermeasure is to raise the PMR to 100\%, which will close the loophole of the practical system. Because the core of the polarization attack is to utilize the difference between the calibrated SNU and the practical one, another countermeasure is the real-time shot noise measurement~\cite{shotnoisem} which is effective for all attacks against the shot noise. To completely close the detection loopholes, the most secure way is the currently proposed continuous-variable measurement-device-independent quantum key distribution protocols~\cite{CVMDI1,CVMDI2,CVMDI3,CVMDI4}, which can remove all known and unknown loopholes on the detection. In the recently proposed CV implementations whose LO is generated "locally"~\cite{LLO1,LLO2,LLO3}, the polarization attack can not be directly implemented to control the practical SNU. But the polarization of the reference pulses could still be controlled by Eve to affect the measurement results, the specific attack against the "local" LO schemes could be analyzed in further research.

\section*{Acknowledgments}

This work was supported by the Key Program of National Natural Science Foundation of China under Grant 61531003, the National Natural Science Foundation under Grant 61427813, 61771439 and 61501414, the National Basic Research Program of China (973 Pro-gram) under Grant 2014CB340102, the China Postdoctoral Science Foundation under Grant 2018M630116 and the Fund of State Key Laboratory of Information Photonics and Optical Communications.

\section*{APPENDIX: Calculation of the secret key rate}
The secret key rate of CV-QKD protocols is calculated with the covariance matrix. We suppose the final quantum state shared by Alice and Bob is ${\rho _{AB}}$ with a covariance matrix

\begin{equation}
{\gamma _{AB}}{\rm{ = }}\left[ {\begin{array}{*{20}{c}}
{\left( {{V_A} + 1} \right){\rm I}}&{\sqrt {T\left( {V_A^2 + 2A} \right)} {\sigma _z}}\\
{\sqrt {T\left( {V_A^2 + 2A} \right)} {\sigma _z}}&{\left( {T{V_A} + 1 + T\epsilon } \right){\rm I}}
\end{array}} \right],
\end{equation}
we only show the secret key rate for reverse reconciliation~\cite{REVERSE}

\begin{equation}
K = \beta I\left( {A:B} \right) - S \left( {E:B} \right),
\end{equation}
where $I\left( {A:B} \right)$ is the classical mutual entropy between Alice and Bob, $S(E:B) = S\left( E \right) - S\left( {E\left| {x_B} \right.} \right)$ is the quantum mutual entropy between Eve and Bob\cite{helevo}, and $\beta$ is the reconciliation efficiency. The mutual information between Alice and Bob reads

\begin{equation}
I\left( {A:B} \right) = \frac{1}{2}{\log _2}\frac{{{V_A}}}{{{V_A} - \frac{{C_{AB}^2}}{{{V_B}}}}}{\rm{ = }}\frac{1}{2}{\log _2}\left( {\frac{{{V_A} + 1 + \chi }}{{\chi  + 1}}} \right),
\end{equation}
where $\chi  = \frac{{1 - T}}{T} + \epsilon $. Since Eve's system E can purify $AB$, we have $S\left( E \right) = S\left( {AB} \right)$. $S\left( {AB} \right)$  is calculated with the two symplectic eigenvalues (${\lambda _1}$, ${\lambda _2}$) of the covariance matrix ${\gamma _{AB}}$~\cite{xma,EXPERIMENT51}

\begin{equation}
S\left( {AB} \right) = G\left[ {{{\left( {{\lambda _1} - 1} \right)} \mathord{\left/
 {\vphantom {{\left( {{\lambda _1} - 1} \right)} 2}} \right.
 \kern-\nulldelimiterspace} 2}} \right] + G\left[ {{{\left( {{\lambda _2} - 1} \right)} \mathord{\left/
 {\vphantom {{\left( {{\lambda _2} - 1} \right)} 2}} \right.
 \kern-\nulldelimiterspace} 2}} \right],
\end{equation}
where $G\left( x \right) = \left( {x + 1} \right){\log _2}\left( {x + 1} \right) - x{\log _2}x$ is the Von Neumann entropy~\cite{von}. Then the system AE is also a pure state after Bob's measurement, $S\left( {E\left| {{x_B}} \right.} \right) = S\left( {A\left| {{x_B}} \right.} \right)$. $S\left( {A\left| {{x_B}} \right.} \right)$ is a function of the symplectic eigenvalue (${\lambda _3}$) of the covariance matrix $\gamma _A^{{x_B}}$

\begin{equation}
S\left( {A\left| {{x_B}} \right.} \right){\rm{ = }}G\left[ {{{\left( {{\lambda _3} - 1} \right)} \mathord{\left/
 {\vphantom {{\left( {{\lambda _3} - 1} \right)} 2}} \right.
 \kern-\nulldelimiterspace} 2}} \right],
\end{equation}
the covariance matrix $\gamma _A^{{x_B}}$~\cite{xma} is

\begin{equation}
\gamma _A^{{x_B}} = {\gamma _B} - C_{AB}^T{\left( {X{\gamma _A}X} \right)^{MP}}{C_{AB}}.
\end{equation}
where $X = diag(1,0)$ and MP denotes the Moore-Penrose inverse of the matrix.

%

\end{document}